\def \SAIT #1 #2 {{\em Mem.\ Soc.\ Astron.\ It.\/} {\bf #1}, #2}
\def \MESS #1 #2 {{\em The Messenger\/} {\bf #1}, #2}
\def \ASTRNACH #1 #2 {{\em Astron. Nach.\/} {\bf #1}, #2}
\def \AAP #1 #2 {{\em Astron. Astrophys.\/} {\bf #1}, #2}
\def \AAL #1 #2 {{\em Astron. Astrophys. Lett.\/} {\bf #1}, L#2}
\def \AAR #1 #2 {{\em Astron. Astrophys. Rev.\/} {\bf #1}, #2}
\def \AAS #1 #2 {{\em Astron. Astrophys. Suppl. Ser.\/} {\bf #1}, #2}
\def \AJ #1 #2 {{\em Astron. J.\/} {\bf #1}, #2}
\def \ANNREV #1 #2 {{\em Ann. Rev. Astron. Astrophys.\/} {\bf #1}, #2}
\def \APJ #1 #2 {{\em Astrophys. J.\/} {\bf #1}, #2}
\def \APJL #1 #2 {{\em Astrophys. J. Lett.\/} {\bf #1}, L#2}
\def \APJS #1 #2 {{\em Astrophys. J. Suppl.\/} {\bf #1}, #2}
\def \APSS #1 #2 {{\em Astrophys. Space Sci.\/} {\bf #1}, #2}
\def \ASR #1 #2 {{\em Adv. Space Res.\/} {\bf #1}, #2}
\def \BAIC #1 #2 {{\em Bull. Astron. Inst. Czechosl.\/} {\bf #1}, #2}
\def \JSQRT #1 #2 {{\em J. Quant. Spectrosc. Radiat. Transfer\/} {\bf #1}, #2}
\def \MN #1 #2 {{\em Mon. Not. R. Astr. Soc.\/} {\bf #1}, #2}
\def \MEM #1 #2 {{\em Mem. R. Astr. Soc.\/} {\bf #1}, #2}
\def \PLR #1 #2 {{\em Phys. Lett. Rev.\/} {\bf #1}, #2}
\def \PASJ #1 #2 {{\em Publ. Astron. Soc. Japan\/} {\bf #1}, #2}
\def \PASP #1 #2 {{\em Publ. Astr. Soc. Pacific\/} {\bf #1}, #2}
\def \NAT #1 #2 {{\em Nature\/} {\bf #1}, #2}

\documentstyle{memsait}
\input epsf.sty
\begin{opening}
\title{DUST ENSHROUDED AGB VARIABLES IN THE LMC} 
\author{PATRICIA WHITELOCK$^1$, MICHAEL FEAST$^2$}
\institute{$^1$South African Astronomical Observatory, PO Box 9,
7935 Observatory, South Africa\\ 
(email: paw@saao.ac.za)\\
$^2$University of Cape Town, 7701 Rondebosch, South Africa\\ 
(email: mwf@artemisia.ast.uct.ac.za)}
\date{} 
\end{opening}

\begin{document}

\oddpagefooter{}{}{} 
\evenpagefooter{}{}{} 
\ 
\bigskip

\begin{abstract} The luminosities and periods of obscured AGB stars in the 
LMC are discussed using a combination of ISO and ground-based infrared
photometry. The bolometric luminosities of these stars fall close to an
extrapolation of the period-luminosity relation derived for oxygen-rich Mira
variables, with both oxygen- and carbon-rich stars falling close to the same
relation.  It has been known for many years that there are, in the
Magellanic Clouds, significant numbers of large amplitude variables which
have considerably higher luminosities than the period-luminosity relation
would predict. Many of these can be shown to be undergoing hot bottom
burning. It is speculated that all large amplitude AGB stars with
luminosities significantly higher than indicated by the period-luminosity
relation are undergoing hot bottom burning.
\end{abstract}

\section{Introduction}
 The asymptotic giant branch (AGB) stars discussed here include those
referred to as Mira variables, OH/IR stars and dusty carbon stars. The
classical defining characteristics of Mira variables are: large amplitudes,
emission line spectra, and periods in excess of about 100 days. The OH/IR
and dusty carbon stars are assumed to be similar to the Miras although they
are often too faint optically to determine visual amplitudes or measure
spectra. Their periods are generally long, up to 1000 days for the carbon
stars and up to about 2000 days for the OH/IR stars. A brief review is given
below of red variables in globular clusters, prior to a more detailed
description of obscured AGB variables in the LMC.

\section{AGB Variables in Globular Clusters}
 Globular clusters provide a well studied and well defined environment in
which to consider AGB variables.  Oxygen-rich Miras are found only in
metal-rich clusters where they are the most luminous stars in the clusters.
The cluster semi-regular (SR) variables are less luminous than the Miras,
but brighter than the non-variable stars.  The luminosities of the Miras are
greater than that of core helium flash, clearly putting them on the AGB.
Broadly speaking stars evolve up the AGB to higher luminosity. But as is now
well known, and as discussed elsewhere in these proceedings an AGB star's
surface luminosity changes quite significantly during the course of the
He-shell flash cycle. It seems likely therefore that stars enter and exit
the Mira phase, possibly more than once, as their luminosity changes during
the course of the shell-flash cycle. There are rather few cluster Miras, as
would be expected of an evolutionary phase that lasts no more than $2 \times
10^5$ years (Renzini \& Greggio 1990). 

In a period-luminosity (PL) diagram we find the low amplitude SR variables
at shorter periods than the Miras. They are on an evolutionary track that
leads to, and terminates at, the Mira PL relation (Whitelock 1986; Feast
1988). Thus we can understand the Mira period luminosity relation as the
locus of the end-points of the evolution of stars with different mass and/or
metallicity. Bedding \& Zijlstra (1998) have recently suggested that the
luminosities of local SR variables (which are presumably younger than the
globular cluster SRs), as determined from Hipparcos parallaxes, follow an
evolutionary track parallel to that found for globular cluster SR variables,
but about 0.8 mag brighter. The SR variables near LMC clusters, studied by
Wood \& Sebo (1996), appear to follow roughly the same track as the local
SRs. At this stage we have little observational evidence for what happens to
stars of even higher initial mass which will evolve at higher luminosity.
If, as mounting evidence suggests, the more massive stars are affected by
hot bottom burning (HBB) it is quite possible that tracks for AGB stars of
several solar masses are quite different.

When the AGB variables drop out of the Mira instability strip, during the 
low luminosity part of the flash cycle, they will
become SR variables for some period of time. This type of SR may show a
detached dust shell, from the previous high mass-loss Mira phase, and/or
abundance anomalies, such as technetium from the dredge-up which accompanies
certain thermal pulses. However, by analogy with the globular
clusters\footnote{In cluster HR diagrams SRs occupy the luminosity range
between non-variables and Miras; unless this represents an evolutionary
sequence there will be an unexplained luminosity gap between the
non-variables and the Miras.} the majority of oxygen-rich SRs must be in a
pre-Mira evolutionary phase and are not expected to show abundance
anomalies.

\begin{figure}
\vspace{7cm}
  \caption[h]{LMC red variables from Wood et al.\ (1992 - fig 8); squares are
supergiants (but see section 4) and circles are AGB stars. Solid symbols are
OH Masers. Fig not included in astro-ph.}
\end{figure}

\section{AGB stars in the Large Magellanic Cloud}

In the following discussion we adopt a distance modulus for the LMC of 18.5
mag, as used in most of the papers to which references are made.  The best
current distance estimates put it slightly further away (e.g. Feast 1999) so
the luminosities discussed here are conservative ones. Our knowledge of AGB
variables in the LMC up to a few years ago is nicely illustrated in the PL
diagram, Fig.~1, taken from Wood et al.\ (1992) which was originally
produced to show the position of the then newly discovered OH/IR stars. The
high luminosity stars were thought to be supergiants, with initial masses in
excess of about $8\,M_{\odot}$. The fainter stars are on the AGB,
and superimposed are model tracks suggesting their approximate initial
masses. The stellar luminosities were derived from single observations of
large amplitude variables, so a good deal of the scatter is due to
variability.  As discussed above, we have good reason to believe that,
certainly at low masses, stars only become Miras for a short while, so that
only part of any evolutionary track is populated by Miras. The picture has
changed somewhat in very recent times with the discovery and detailed
investigation of more AGB variables with thick shells, as discussed below.

Feast et al.\ (1989) established PL relations for carbon-
and oxygen-rich Miras in the LMC. These were derived from large amplitude
variables monitored over their light cycles to derive mean luminosities. At
$K$ the carbon and oxygen stars obey the same PL relation with a scatter of
only 0.13 mag, up to periods of around 420 days. Bolometric luminosities
were determined by fitting blackbodies to $JHK$ photometry - a reasonable
approximation for these stars which have only very thin dust shells. The
bolometric PL relations appeared to be different for the oxygen- and
carbon-rich Miras, although it was never clear if this was really so, or if
it was an artifact of the way the bolometric luminosities were derived.
Groenewegen \& Whitelock (1996) derived a bolometric PL relation for LMC
carbon Miras using all the available data for spectroscopically confirmed C
stars, although there were only single epoch observations for many of these.
Their PL relation was essentially indistinguishable from that for the O-rich
stars. At the time this work was done no LMC carbon Miras were known with
periods significantly longer than 500 days, and the optically visible
oxygen-rich stars with periods over 420 days lay clearly above the PL
relationship. One of these luminous AGB stars, RCG\,69 (0523--6644), was
among the sample looked at by Smith et al.\ (1995) for lithium, and is in
fact lithium-rich.

Smith et al.\ (1995) made a survey for lithium among AGB stars in the SMC
and LMC, and found that a very large fraction of those with 
$-6 > M_{\it bol} > -7$ were lithium rich, as were a much smaller number of
lower luminosity stars. Our present understanding of lithium enhancements in
AGB stars (e.g. Sackmann \& Boothroyd 1992) is that they occur principally
as a result of HBB. Towards the end of AGB evolution, for stars with an
initial mass in the range 4 to 6 $M_{\odot}$, the base of the convective
envelope can dip into the H-burning shell, with far reaching consequences
for the evolution of the star. The transition from O- to C-rich is affected;
exactly how seems to depend on the model and in particular on how mass loss
is treated. HBB may prevent carbon stars forming at all, or prevent them
from happening until the envelope mass is depleted (Frost et al.\ 1998).  In
some models C stars do form and then HBB turns them back into O-rich stars
(Marigo et al.\ 1999).  Another consequence, for stars in a rather narrow
mass range, is the formation of lithium via beryllium.  One of the most
important results in the present context is a rapid increase in luminosity
(Bl\"ocker \& Sch\"onberner 1991); again the details depend on mass loss
which tends to decrease the effect of HBB.  In view of this it is
interesting to note that most of the stars illustrated in fig.~6 of Smith at
al.\ (1995) lie above the PL relation and many of these must be undergoing
HBB, as their lithium abundance indicates. As it is possible for some stars
to experience HBB without showing lithium enhancements (e.g. Mazzitelli et
al. 1999) we speculate that all of the stars with luminosities above the PL
are there because of HBB. Note that the group of stars discussed by Smith et
al.\ is not representative of AGB stars generally.  These are the stars for
which it is practical to get optical spectra with sufficient resolution and
signal-to-noise to measure lithium line strengths, i.e. the brightest ones
with the thinnest dust shells. Furthermore, the luminosities of these stars
require more detailed investigation; most are based on single observations
of large amplitude variables. The periods are not all well defined and it is
possible that some of the stars are SR variables and should not properly
be included in this discussion.

Wood (1998) discussed the results of a long term investigation of nine
obscured IRAS sources in the LMC, providing periods ($\rm 530 <P< 1295$
days) and luminosities for them. Six of these sources are significantly
fainter than an extrapolation of the PL relation and he speculated that
these were C stars - available evidence on the subject being somewhat
contradictory. IR spectroscopy, either from the ground or from ISO ({\it
ISOPHOT} or {\it ISOCAM}), demonstrates that all but one of Wood's nine
sources are indeed C stars, the presence of the $\rm 3\,\mu m$ $\rm
C_2H_2+HCN$ feature being a clear diagnostic of carbon-rich chemistry (van
Loon et al.\ 1999b or references therein). The one exception was the star
with the longest period which Wood et al.\ (1992) had already shown to be an
OH source, and therefore O rich. Wood's (1998) conclusion, from the
discovery of apparently subluminous C stars, was that his observations
``{\it $\cdots$ clearly demonstrate that once significant mass loss and dust
formation occurs, large amplitude LPVs no longer fall on the tight
$M_{bol}/\log P$ relation found for Mira variables with $P<450$ days}".

\begin{figure}
\epsfxsize=7.7cm 
\hspace{9cm}\epsfbox{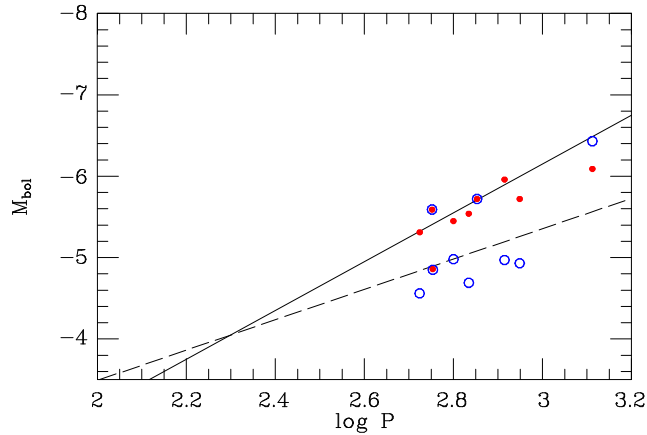} 
  \caption[h]{A comparison of luminosities from Wood (1998) (open circles) 
and van Loon et al.\ (1999b) (closed circles) for the stars in common. The
solid and broken lines are extrapolations of the PL relations for O- and
C-rich stars, respectively, determined by Feast et al.\ (1989).}
\end{figure}

The nine stars discussed by Wood are also a subgroup of about 50 LMC
sources, originally selected from IRAS data, for which a group of us have
been obtaining ground-based and ISO data over the last few years (van Loon
et al.\ 1999b and references therein). We have derived periods for them,
which can be further refined by combining our data with Wood's; these do not
differ significantly from those measured by Wood.  Van Loon et al.\ (1999b)
combined {\it ISOCAM} and/or {\it ISOPHOT} photometry and spectroscopy with
ground based, $JHKL$, photometry and fitted models to derive an independent
luminosity for these (and other) stars. These results illustrate the
luminosity {\it at the time of the ISO observations}, i.e. at random phase,
whereas Wood (1998) combined IRAS observations, after applying a correction
intended to reproduce mean light, with mean near-infrared photometry. A
comparison of our estimates of the C-star luminosities with those from Wood
is shown in Fig.~2. Three of the stars agree, while we find brighter
luminosities for the other five. The OH/IR star is slightly fainter than
Wood found. Our observations were made at random phases, and a rough check
suggests that two are close to mean light, two fainter and four brighter
than mean light. The reason for the apparent systematic difference between
the van Loon at al.\ and the Wood luminosities is not immediately obvious
and requires more detailed investigation, but in view of Fig.~2 and the
discussion below it seems premature to conclude that sources with thick dust
shells fall systematically below the PL relation.

The analysis of our complete sample of LMC stars is at a preliminary stage
so the details of the following discussion may change. Combining preliminary
periods with luminosities derived by van Loon et al.\ (1999b), in the same
way as those discussed for the Wood (1998) sources, we obtain the PL data
illustrated in Fig.~3. Several of the long period O-rich stars are also OH
sources (Wood et al.\ 1992). Extrapolations of the Feast et al.\ (1989) PL
relations are shown; note that the PL relation determined by Groenewegen \&
Whitelock (1996) for C stars extrapolates between the two lines illustrated.
In interpreting Fig.~3 it is crucial to remember that the luminosities shown
here are one-off measurements of large amplitude variables, because we have,
in general, only single epoch ISO observations. It is therefore obvious that
a good deal of the scatter in this diagram is due to variability and
multiple observations of three stars give some impression of this. It is
also clear that these data scatter around the extrapolated O-rich PL
relation, with both C- and O-rich stars falling close to the same line.

\begin{figure}
\epsfxsize=7.7cm 
\hspace{8.5cm}\epsfbox{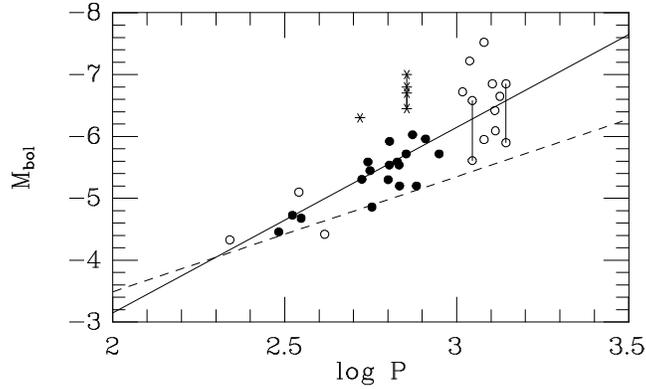} 
  \caption[h]{The PL relation for long period AGB stars in the LMC.
The luminosities are from van Loon et al.\ (1999b) and are single phase
measurements. The periods are from Whitelock et al.\ (in preparation). The
solid and broken lines are extrapolations of the PL relations, determined by
Feast et al.\ (1989), for O- and C-rich stars, respectively. Solid symbols
represent C stars and open ones O-rich stars. The stars marked as asterisks
are IRAS\,04496--6958 and SHV\,F4488 which are both C-rich (but see text).
Connected points represent measurements of the same star at different
epochs.}
\end{figure}

The best sampled light curves for the LMC carbon stars have data spanning 5
to 6 years and show peak-to-peak amplitudes of $\Delta K \sim 2$ mag.
Several of the C stars exhibit apparently secular variations on top of the
regular pulsations, as do many galactic C-stars with high mass-loss rates
(Whitelock et al.\ 1997), probably the results of fluctuations in the
mass-loss rate. It is difficult to establish the bolometric amplitudes of
these stars as few have been monitored at wavelengths longer than $3 \rm \mu
m$ where most of the energy is emitted. Nevertheless, the discussion by van
Loon at al.\ (1998) suggests that the C stars may have $\Delta m_{bol}
\sim 1.0$, while the O-rich stars with $P>1000$ days will have $1.0< \Delta
m_{bol} < 1.5$ mag.

\section{HBB in Individual Stars}
 
The four separate estimates of the luminosity  of IRAS\,04496--6958 are
shown as connected asterisks in Fig.~3 (van Loon et al.\ 1998, 1999a,
1999b). The $\rm 3\, \mu m$ spectrum of this star shows a clear $C_2H_2+HCN$
absorption feature, indicating it is a C star. Its $\rm 10\, \mu m$ spectrum
(Fig.~4), however, is very unusual in that it shows both silicate and
silicon carbide features, suggesting a mixed O- and C-rich chemistry (Trams
et al.\ 1999). This is the first example of an extragalactic carbon star
with silicate emission. There are several galactic stars which show similar
combination features; they are generally understood to be binary systems in
which one component is a carbon star and the silicate dust resides in a
circum-binary disk (Lloyd Evans 1990). It may be that the same explanation
applies here, although in view of its high luminosity, Trams et al.\ (1999)
offered an alternative explanation -- that IRAS\,04496--6958 was, until recently,
a star undergoing HBB, hence the silicate dust. It would thus be an example
of a star in which HBB terminated, perhaps due to mass-loss, and which then
underwent a thermal pulse and dredge-up turning it into a C star.

\begin{figure}
\epsfysize=6.8cm 
\hspace{3.5cm}\epsfbox{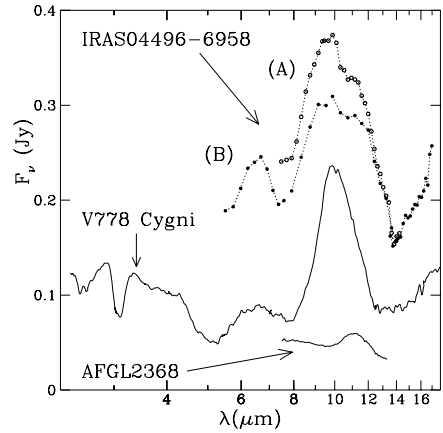} 
  \caption[h]{{\it ISOCAM} spectrum of IRAS\,04496--6958 exhibiting
both silicate and SiC emission features. Shown for comparison are V778\,Cyg,
a C star with a silicate shell, and AFGL\,2368, a thick shelled C star.
Figure from Trams et al.\ (1999)}
\end{figure}

The solitary asterisk in Fig.~3 represents a C star which Smith et al.\
(1995) found to be lithium rich, SHV\,F4488. The luminosity used here,
$M_{bol}=-6.3$, is considerably brighter than that quoted by Smith et al.,
$M_{bol}=-5.7$, which was taken from Hughes \& Wood (1990). The difference
could be due to variability or possibly Hughes \& Wood, who had 
$JHK$ photometry from only one epoch, underestimated the flux. In any case
the enhanced lithium tells us that the star is undergoing HBB.

IRAS\,04496--6958 and SHV\,F4488 are more luminous than the other C stars
shown in Fig.~3. They lie in the same region of the PL diagram, above the
extrapolation of the Mira relation, as do the luminous O-rich AGB stars
discussed by Feast et al.\ (1989) and as do almost all of the lithium-rich
stars discussed by Smith et al.\ (1995).

Finally, it should be noted that one of the stars marked as a short
period supergiant in Fig.~1, HV\,2572, is among the Smith et al. (1995)
lithium-rich sample and must therefore be an AGB star undergoing HBB,
not a supergiant. 

\section{Conclusions}

Observations of large amplitude variables
in the LMC show the following: \begin{enumerate}
\item those which are close to the end of their AGB  lifetimes fall close to
the Mira PL relation;
\item C- and O-rich Miras obey the same PL as well as we can currently 
establish;
\item many, perhaps all, of the AGB variables which lie above the PL relation 
are undergoing HBB. 
\end{enumerate}

There is a caveat on item 2 above: we cannot as yet eliminate the
possibility that there are stars which lie below the PL, as limitations in
the sensitivity of our surveys may have prevented us from detecting them as
yet. In this regard we should look at the Galactic Centre where there is
evidence for long-period large-amplitude variables with low luminosities
(Blommaert et al.\ 1998; Wood et al. 1998). Finally, if we really want to
know the luminosities of these large-amplitude variables we need to monitor
them at longer wavelengths than has been done to date.


\acknowledgements We are grateful to our colleagues, particularly Jacco
van Loon and Albert Zijlstra, for allowing us to discuss data in advance of
publication. We also thank Jacco van Loon and John Menzies for a critical
reading of a draft of this manuscript.

\end{document}